\documentclass[twocolumn,english,pra,aps,superscriptaddress]{revtex4-2}
\usepackage[T1]{fontenc}
\usepackage{color}
\usepackage{babel}
\usepackage{mathtools}
\usepackage{bm}
\usepackage{graphicx}
\usepackage{esint}
\usepackage{amsfonts}
\usepackage{graphicx}
\usepackage{float}
\usepackage{subfigure}
\usepackage{amsthm}
\usepackage{amssymb}
\usepackage{braket}
\usepackage[usenames,divipsnames,svgnames,table]{xcolor}
\usepackage[unicode=true,pdfusetitle,
bookmarks=true,bookmarksnumbered=false,bookmarksopen=false,
breaklinks=true,pdfborder={0 0 0},backref=false,colorlinks=true]{hyperref}
\hypersetup{linkcolor=NavyBlue,urlcolor=NavyBlue,citecolor=NavyBlue}
\usepackage{breakurl}

\newtheorem{theorem}{Theorem}

\usepackage{orcidlink}

\begin{document}

\title{Evaluation of quantum Fisher information for large systems}

\author{Qi Liu\orcidlink{0000-0002-5716-6461}}
\email{d202280037@hust.edu.cn}
\affiliation{National Precise Gravity Measurement Facility, MOE Key
Laboratory of Fundamental Physical Quantities Measurement, School of Physics,
Huazhong University of Science and Technology, Wuhan 430074, China}

\begin{abstract}
Quantum Fisher information (QFI) plays a vital role in quantum precision measurement, quantum information, many-body physics, and other domains.
Obtaining the QFI from experiment for a quantum state reveals insights such as the limits of estimation accuracy for a certain parameter, the degree of entanglement, and the geometric characteristics of the quantum state.
Nonetheless, the measurement complexity of the QFI and its lower bound hinges on the dimension of the quantum state. 
Consequently, reducing the complexity of measurement is a significant challenge.
This paper presents a methodology for evaluating the QFI of high-dimensional systems by transferring information to an auxiliary system and measuring its sub-QFI, while also offering conditions to diminish the dimension of auxiliary system to be measured without affecting the amount of information obtained by it.
\end{abstract}

\maketitle

\section{Introduction}
The quantum Fisher information (QFI), which encapsulates the information related to quantum states, has applications in various domains such as quantum information and quantum precision measurement.
For instance, in the quantum Cram\'{e}r-Rao bound\cite{Helstrom1976,Holevo1982}, the derivative of the QFI can characterize the precision limit of a specific quantum state for parameter estimation.
Furthermore, it can act as an entanglement witness\cite{Toth2012}, enabling the assessment of the entanglement degree of $N$ qubit quantum state, specifically indicating the number of qubits entangled with one another. 
QFI is also associated with quantum geometric tensors (QGT). 
The QGT is defined as
\begin{equation}
Q_{\mu\nu}=\braket{\partial_{\mu}\psi|\partial_{\nu}\psi}-\braket{\partial_{\mu}\psi|\psi}\braket{\psi|\partial_{\nu}\psi},
\end{equation}
where the real part is proportional to QFI, while the imaginary part is referred to as the Berry curvature. 
The QGT finds utility in the study of many-body physics, including phenomena such as quantum phase transition\cite{Venuti2007,Zanardi2007} and the quantum Hall effect\cite{TOzawa2018,Bauer2016}. 

Consequently, by obtaining the QFI of the quantum state, one can derive information regarding the estimation accuracy limit of the parameters containing in quantum state, the degree of entanglement and the geometric properties of the quantum state. 
Quantum state tomography (QST) is the most straightforward approach for determining the QFI of a quantum state.
However, QST presents significant measurement complexity\cite{Haah2017}.
Therefore, various methods have been developed for experimental measurement of the QFI, including periodic drive\cite{Ding2022,Tan2019}, sudden quench\cite{Ozawa2018,Yu2020} and Loschmidt echo\cite{Macri2016}.

However, direct measurement of the QFI poses significant challenges, necessitating a restricted change in parameter.
For instance, in the period-driven scheme, parameter must vary according to a specific period to measure the population of the quantum state, with the population needing to be integrated with over the period to derive the QFI value.
Consequently, some researchers choose to measure the lower bound of QFI, specifically the sub-quantum Fisher information (sub-QFI). 
The sub-QFI arises from an upper bound of fidelity~\cite{Miszczak2009,Jacob2022}, 
represented as $\sqrt{g(\rho_{1},\rho_{2})}\ge f(\rho_{1},\rho_{2})$.
The function $f(\rho_1,\rho_2)=\mathrm{Tr}\left(\sqrt{\rho_{1}\sqrt{\rho_{2}}\rho_{2}}\right)$ is referred to as Uhlmann's quantum fidelity~\cite{Hayashi2004,Braunstein1994}, while $g(\rho_{1},\rho_{2})$ is called superfidelity, which can be expressed as
\begin{equation}
g(\rho_{1},\rho_{2})=\mathrm{Tr}(\rho_{1}\rho_{2})
+\sqrt{\left(1-\mathrm{Tr}(\rho_{1}^{2})\right)\left(1-\mathrm{Tr}(\rho_{2}^{2})\right)}.
\end{equation}
Analogous to the relationship between fidelity and QFI, the negative value of the coefficient of the second-order term of $g(\rho(x),\rho(x+\mathrm{d}x))$ is proportional to the sub-QFI, specifically
\begin{equation}
F^{(\mathrm{sub})}=8\lim_{\mathrm{d} x\to0}\frac{1-\sqrt{R(\rho(x),\rho(x+\mathrm{d}x))}}{\mathrm{d}x^{2}}.
\end{equation}
From the aforementioned equation, it is evident that sub-QFI serves as a lower bound for QFI.
The sub-QFI possesses properties analogous to those of the QFI, and for the quantum state parameterized by unitary processes $e^{-ixH}$ can be expressed as~\cite{Cerezo2021}
\begin{equation}
F^{\left(\mathrm{sub}\right)}=\mathrm{Tr}(\rho^{2}H^{2})-\mathrm{Tr}(\rho H\rho H).
\end{equation}
In Appendix \ref{arbitrary sub-QFI}, we demonstrate the form of sub-QFI for mixed states, which is independent of the processes of parameterization, is
\begin{equation}
F^{\left(\mathrm{sub}\right)}=2\mathrm{Tr}\left[(\partial_{x}\rho)^{2}\right]
+\frac{1}{2}\frac{\left[\partial_{x}\mathrm{Tr}(\rho^{2})\right]^{2}}{1-\mathrm{Tr}(\rho^{2})}.
\label{general sub-QFI}
\end{equation}
When the quantum state is pure, the sub-QFI is equivalent to its QFI, expressed as $F^{(\mathrm{sub})}=2\mathrm{Tr}\left[\left(\partial_{x}\rho\right)^{2}\right]$. 
Furthermore, for a two-dimensional system, its superfidelity $g(\rho(x),\rho(x+\mathrm{d}x)$ is precisely equal to the fidelity value $f(\rho(x),\rho(x+\mathrm{d}x)$, indicating that its sub-QFI is also equivalent to its QFI.

The sub-QFI can be measured experimentally, enabling the evaluation of the QFI based on the measurement results. 
The primary step in obtaining the sub-QFI involves measuring the value of $g(\rho(x),\rho(x+\mathrm{d}x)$. Once the superfidelity is determined, it can be utilized to calculate the sub-QFI of state $\rho^{b}$ using equation (3).
The objective of measuring superfidelity is to derive the value of $\mathrm{Tr}\left(\rho_{1}\rho_{2}\right)$, which can be accomplished through swap test\cite{Ekert2002} and randomized measurements\cite{Rath2021,Yu2021}. 
However, the measurement complexity is contingent upon the dimension of the system being measured. In the case of swap test, quantum gates must be prepared to accommodate the quantum state, while for randomized measurement, the number of measurements is approximately $2^{aN}$, $N$ denotes the dimension of quantum state and $a\approx1$\cite{Elben2018,Elben2020,Vermersch2019}.
It is evident that as the dimension of the system increases, the complexity of measuring sub-QFI also escalates.
In light of this, we propose a scheme to evaluate the QFI of the initial state by introducing an auxiliary system and measuring the sub-QFI associated with this auxiliary system. 
We aim to reduce measurement complexity by minimizing the dimension of the auxiliary system as much as feasible.

\section{Methodology}

\begin{figure}[H]
\centering
\includegraphics[scale=0.27]{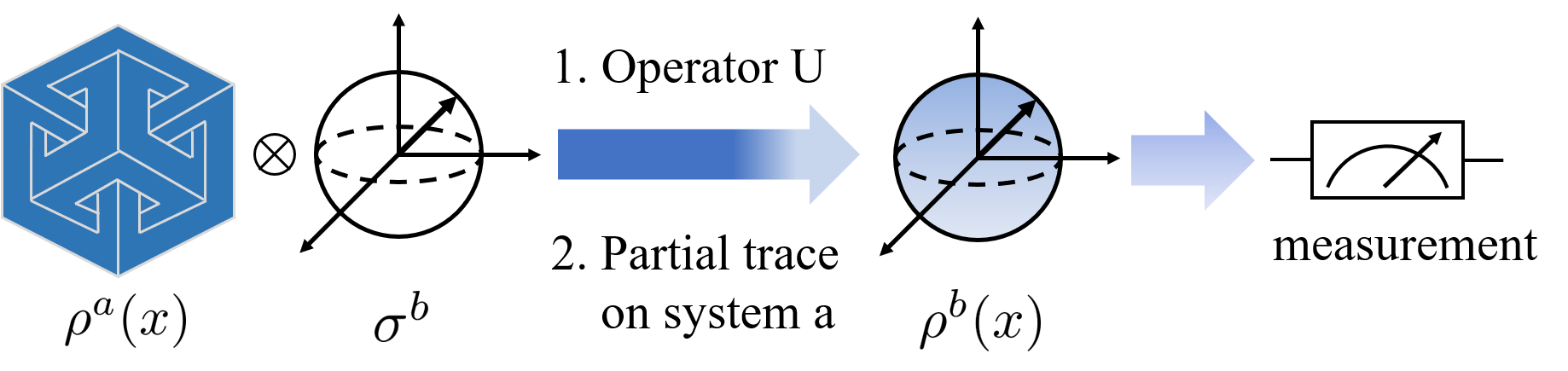}
\caption{The process of evaluating the QFI of the initial quantum state using the auxiliary system. $\rho^{a}(x)$ is the initial quantum state and depends on the parameter. $\sigma^{b}$ is the quantum state in auxiliary system, which is initially independent of parameter. They form a composite system. By performing a unitary transformation on it and taking the partial trace on the system $a$, one can obtain a parameter-dependent quantum state $\rho^{b}(x)$ in the auxiliary system. Finally, the value of its sub-QFI is measured to evaluate the QFI of initial state $\rho^{a}(x)$}
\label{fig:flowchart}
\end{figure}

The central tenet of our approach is to propagate the information about the parameter in the initial system to an auxiliary system, and eventually measure the sub-QFI of this auxiliary system to evaluate the QFI of the initial system. 
The flowchart illustrating the scheme is presented in Figure \ref{fig:flowchart}.

Specifically, let us consider the $N$-dimensional initial state $\rho^{a}(x)$ in system $a$,
which is dependent on parameter $x$.
In the initial phase, we introduce a state $\sigma^{b}$ in the auxiliary system $b$, 
which is independent of parameter, combining it with $\rho^{a}(x)$ to form the composite state $\rho^{ab}(x)=\rho^{a}(x)\otimes\sigma^{b}$.
Utilizing the property of QFI, that is $F(\rho^{a}(x)\otimes\sigma^{b})=F(\rho^{a}(x))+F(\sigma^{b})$~\cite{Toth2014}, we equate the QFI of the composite state to that of the initial state. 
Subsequently, we apply a unitary transformation to the composite state, adopting a design inspired by the work in Ref.~\cite{Lu2013}, with the unitary operator $U$ defined as
\begin{equation}
U=\sum_{i=1}^{N}\Pi_{i}^{a}\otimes O_{i}^{b}.
\end{equation}
$\{\Pi_{i}^{a}\}$ are projectors in system $a$, and $\{O_{i}^{b}\}$ are unitary operators in system $b$.
The $N$ unitary operators must not be identical, because in this way, the information in $\rho^{a}(x)$ cannot propagate into the auxiliary system.
The QFI of composite state after unitary transformation is invariant, that is $F(\rho^{ab})=F(U\rho^{ab}U^{\dagger})$.
Next, we perform partial trace on system $a$, the obtained state is
\begin{equation}
\mathrm{Tr}_{a}\left[U(\rho^{a}\otimes\sigma^{b})U^{\dagger}\right]=\sum_{i=1}^{N}\mathrm{Tr}\left(\Pi_{i}^{a}\rho^{a}\Pi_{i}^{a}\right)O_{i}^{b}\sigma^{b}O_{i}^{b\dagger}.
\end{equation}
For partial trace, the inequality is observed $F(U\rho^{ab}U^{\dagger})\ge F\left[\mathrm{Tr}_{a}(U\rho^{ab}U^{\dagger})\right]$. 
The coefficients $\{\mathrm{Tr}\left(\Pi_{i}^{a}\rho^{a}\Pi_{i}^{a}\right)\}$ depend on parameters, that is, some information of parameter have been transferred from initial system $a$ to auxiliary system $b$.
Ultimately, we can measure the superfidelity of $\rho^{b}(x)$ and $\rho^{b}(x+\mathrm{d}x)$ and use it to obtain the value of sub-QFI.
The inequality $F(\rho^{b})\ge F^{(\mathrm{sub})}(\rho^{b})$ maintained.
Thus, by directly measuring the superfidelity of state $\rho^{b}$ in auxiliary system, we can get a lower bound of QFI of initial state.

To sum up, the above process is encapsulated by the inequality
\begin{equation}
F(\rho^{a})=F\left(U\rho^{ab}U^{\dagger}\right)\ge
F\left[\mathrm{Tr}_{a}\left(U\rho^{ab}U^{\dagger}\right)\right]
\ge F^{(\mathrm{sub})}\left(\rho^{b}\right).\label{scheme inequality}
\end{equation}
To approximate the sub-QFI of the auxiliary system to that of the initial system, the information about the parameter in the initial system should be transferred to the auxiliary system as much as possible.
This is achieved by maximizing $F\left[\mathrm{Tr}_{a}\left(U\rho^{ab}U^{\dagger}\right)\right]$, a goal accomplished through the optimization of the auxiliary system $\sigma^{b}$ and the unitary transformation $U$.
Moreover, in order to reduce the complexity of measuring sub-QFI, the dimension of the auxiliary system should be reduced.
Therefore, the appropriate $\sigma^{b}$ and $U$ should aim to minimize the dimension of the auxiliary system while maintaining the maximum value of $F\left[\mathrm{Tr}_{a}\left(U\rho^{ab}U^{\dagger}\right)\right]$. 

In Section \ref{single qubit}., we initially focus on optimizing the case that the auxiliary system is a single qubit, and elaborate the measurement scheme and optimization process.
In Section \ref{arbitary}., we broaden the scope of optimization to the case of quantum state with arbitrary dimension, and give the maximum $F\left[\mathrm{Tr}_{a}\left(U\rho^{ab}U^{\dagger}\right)\right]$ and the condition of reducing the dimension of the auxiliary system without changing the maximum value.

\section{Single qubit auxiliary system} \label{single qubit}
Assuming the quantum state of the auxiliary system is a single qubit.
Following the aforementioned method, we define the unitary operator applied to the composite state as $U=\Pi_{1}^{a}\otimes O_{1}^{b}+\Pi_{1}^{a}\otimes O_{2}^{b}$, yielding the resulting state after partial trace as
\begin{equation}
\rho^{b}=\mathrm{Tr}\left(\Pi_{1}^{a}\rho^{a}\Pi_{1}^{a}\right)O_{1}^{b}\sigma^{b}O_{1}^{b\dagger}+\mathrm{Tr}\left(\Pi_{2}^{a}\rho^{a}\Pi_{2}^{a}\right)O_{2}^{b}\sigma^{b}O_{2}^{b\dagger}.
\end{equation}
To simplify notation, we denote the trace $\mathrm{Tr}(\Pi_{i}^{a}\rho^{a}\Pi_{i}^{a})$ as $p_{i}$ and represent the operator $O_{i}^{b}\sigma^{b}O_{i}^{b\dagger}$ as $\rho_{i}$.
The coefficients $p_1$ and $p_2$ depend on parameter, while the density operator $\rho_1$ and $\rho_2$ do not.
Within the Bloch sphere representation, the density operator $\rho^{b}$ is characterized by
\begin{equation}
\rho^{b}=\frac{1}{2}\left[\mathbb{I}+(p_{1}\vec{r}_{1}+p_{2}\vec{r}_{2})\cdot\vec{\sigma}\right].
\end{equation}
The Bloch vector $\vec{r}$ of $\rho^{b}$, as evident from the equation, is the sum $p_{1}\vec{r}_{1}+p_{2}\vec{r}_{2}$.
The function $\chi(\theta,\phi)$ is defined as
$\chi(\theta,\phi)=\sin\theta_{1}\sin\theta_{2}\cos(\phi_{1}-\phi_{2})+\cos\theta_{1}\cos\theta_{2}-1$, where $\theta_{1}(\theta_{2})$ and $\phi_{1}(\phi_{2})$ represent the polar angle and azimuth angles of Bloch vector $\vec{r}_{1}\left(\vec{r}_{2}\right)$, respectively.
The QFI of $\rho^{b}$ can be calculated as
\begin{equation}
F(\rho^{b})=\frac{\left[2r^{2}(r^{2}-1)\chi(\theta,\phi)+r^{4}\chi^{2}(\theta,\phi)\right](\partial_{x}p_{1})^{2}}
{1-r^{2}-2r^{2}p_{1}p_{2}\chi(\theta,\phi)}.
\end{equation}
In the above equation, $r=\left|\vec{r}_{1}\right|=\left|\vec{r}_{2}\right|$. Through optimization in appendix \ref{opt single-qubit}, it can be obtained that the maximal QFI of $\rho^{b}$ is 
\begin{equation}
\max F(\rho^{b})=\frac{(\partial_{x}p_{1})^{2}}{p_{1}p_{2}},
\end{equation}
this is achievable when $\chi(\theta,\phi)=-2$ and $r=1$.
The condition $r=1$ indicates the auxiliary $\sigma^{b}$ is a pure state, and $\chi(\theta,\phi)=-2$ signifies $\rho_{1}$ and $\rho_{2}$ are orthogonal.

In summary, the optimal state for a two-dimensional auxiliary system is a single qubit pure state.
To ensure the QFI of $\rho^{b}$ attains its maximal value, the operators $O_{1}^{b}$ and $O_{2}^{b}$ within $U$ must be orthogonal.
Additionally, we observe that the coefficients take the form $p_{1}=\mathrm{Tr}\left(\Pi_{1}^{a}\rho^{a}\Pi_{1}^{a}\right)$ and $p_{2}=\mathrm{Tr}\left(\Pi_{2}^{a}\rho^{a}\Pi_{2}^{a}\right)$, indicating the possibility of outcomes for  measurements $\Pi_{1}^{a}$ and $\Pi_{2}^{a}$ on $\rho^{a}$. 
Consequently, the maximal QFI of $\rho^{b}$ corresponds to the classical Fisher information (CFI)~\cite{Fisher1922,Fisher1925} for $\rho^{a}$ under the projections $\{\Pi_{1}^{a},\Pi_{2}^{a}\}$.
These projections must constitute the optimal measurements for $\rho^{a}$ to guarantee equivalence in QFI between $\rho^{a}$ and $\rho^{b}$.
For a single qubit, the sub-QFI is equivalent to the QFI.
By identifying the optimal measurements $\{\Pi_{1}^{a},\Pi_{2}^{a}\}$ for $\rho^{a}$, from above method,
we can achieve $F\left(\rho^{a}\right)=F\left(\rho^{b}\right)=F^{(\mathrm{sub})}\left(\rho^{b}\right)$.
Thereby,  a direct measurement toward the QFI for $\rho^{a}$ is achievable.

To elucidate the above scheme's efficacy in simplifying the complexity of the measurement, we consider an example of that assesses two-qubit entanglement through measuring its QFI.
As mentioned above, QFI can be used as the entanglement witness.
If a $N$-qubit quantum state violates the inequality
\begin{equation}
F\left[\rho,J_l\right]\le \left(N-1\right)^{2}+1,
\end{equation}
this state is genuine multipartite entangled.
$F\left[\rho,J_l\right]$ is the QFI of
$e^{-ix J_l}\rho e^{ix J_l}$, $J_l=\sum_{k}\sigma_{l}^{(k)}/2$,
$l=x,y,z$, and $k$ is the index of different qubit.
For a Bell state $\ket{\psi}=(\ket{00}+\ket{11})/\sqrt{2}$~\cite{Bell1987}, 
\begin{equation}
\ket{\psi_{x}}=e^{-ix J_{z}}\ket{\psi}=\frac{1}{\sqrt{2}}\left(e^{-ix}\ket{00}+e^{ix}\ket{11}\right).
\end{equation}
The QFI $F\left[\ket{\psi_{x}},J_z\right]$ of $\ket{\psi_{x}}$ with parameter $x$ can be calculated using
\begin{equation}
F=4\left(\braket{\partial_{x}\psi_{x}|\partial_{x}\psi_{x}}
-\left|\braket{\psi_{x}|\partial_{x}\psi_{x}}\right|^{2}\right).
\end{equation}
Thus, for Bell state, the QFI $F[\ket{\psi_{x}},J_{z}]=4$,
which can be used to determine that this quantum state is genuine entangled.

Using the above methodology, we introduce a single qubit $\sigma^{b}=\ket{0}\bra{0}$ as the auxiliary system to form a composite system with $\rho^{a}$.
The unitary operator acts on the composite system is designed as
\begin{equation}
\begin{aligned}
U&= \frac{1}{2}\left(\ket{00}+\ket{11}\right)\left(\bra{00}+\bra{11}\right)\otimes \mathbb{I}\\
&+\frac{1}{2}\left(\ket{00}-\ket{11}\right)\left(\bra{00}-\bra{11}\right)\otimes X.
\end{aligned}
\end{equation}
The $X$ gate, denoted by $X$, effects a flip of the single qubit.
Then, the state $\rho^{b}$ after partial trace is
\begin{equation}
\rho^{b}=\frac{1}{2}\left(1+\cos2x\right)\ket{0}\bra{0}+\frac{1}{2}\left(1-\cos2x\right)\ket{1}\bra{1}.
\end{equation}
Evidently, the information of the parameter $x$ is encoded within the quantum state $\rho^{b}$ in auxiliary system, and the QFI of $\rho^{a}$ matches that of $\ket{\psi_{x}}$.
Consequently, the selection of suitable projection operators $\{\Pi_{1}^{a},\Pi_{2}^{a}\}$ ensures complete information transfer from $\ket{\psi_{x}}$ to $\rho^{b}$.
And since the auxiliary system is single qubit, its sub-QFI is also equal to the value of its QFI, that is, sub-QFI of $\rho^{b}$ is $F^{(\mathrm{sub})}(\rho^{b})=F(\rho^{b})=F\left[\ket{\psi_{x}},J_{z}\right]=4$. 
Thus, by measuring the sub-QFI of quantum state $\rho^{b}$ in auxiliary system, one can obtain the exact value of QFI of $\rho^{a}$ and determine whether this quantum state is entangled or not.

Furthermore, as measurement complexity for the sub-QFI grows with the system's dimensional, simplifying the measurement from a two-qubit to a single-qubit scenario from the above method effectively reduces this complexity, fulfilling our objective.

\section{Optimize the method} \label{arbitary}
Through the single qubit example above, it is evident that the auxiliary system $\sigma^{b}$, 
projection operators $\{\Pi_{i}^{b}\}$ and unitary operators $\{O_{i}^{b}\}$ in Eq.\ref{scheme inequality}, can be optimized to maximize the QFI of $\rho^{b}(x)$. 
We proceed to discuss the optimization of quantum states for arbitrary dimension. To do this, we need to use the following theorem.

\begin{theorem} \label{theorem1}
Suppose a quantum state consists of a set of linearly independent quantum states $\{\ket{\psi_{i}}\}$ with form $\rho(x)=\sum_{i}p_{i}\ket{\psi_{i}}\bra{\psi_{i}}$, where the coefficients $\{p_{i}\}$ depend on the parameter, and $\{\ket{\psi_{i}}\}$ do not.
The QFI of quantum state with this form is 
\begin{equation}
F=\sum_{i=1}^{N}\frac{\left(\partial_{x}p_{i}\right)^{2}}{p_{i}}
-\frac{1}{2}\sum_{i,j=1}^{N}S_{ij}\left(\frac{\partial_{x}p_{i}}{p_{i}}-\frac{\partial_{x}p_{j}}{p_{j}}\right)\left(SL\right)_{ij}.
\end{equation}
$S_{ij}=\braket{\psi_{i}|\psi_{j}}$ is the element of Gram matrix and $L_{ij}$ is the element of SLD operator under bases $\{\ket{\psi_{i}}\}$.
The maximum value of this QFI is
\begin{equation}
F_{\mathrm{max}}=\sum_{i=1}^{N}\frac{\left(\partial_{x}p_{i}\right)^{2}}{p_{i}},
\end{equation}
which can be achieved by letting $S_{ij}=\delta_{ij}$, that is the states in $\{\ket{\psi_{i}}\}$ are orthogonal to each other.
\end{theorem}

The proof of this theorem is provided in Appendix \ref{orthogonal bases}.
In the case where the quantum state in auxiliary system is pure, the state after partial trace $\rho^{b}$ can be written as
\begin{equation}
\rho^{b}=\sum_{i=1}^{N}\mathrm{Tr}\left(\Pi_{i}^{a}\rho^{a}\Pi_{i}^{a}\right)\ket{\psi_{i}}\bra{\psi_{i}}.
\end{equation}
Denoting $\mathrm{Tr}\left(\Pi_{i}^{a}\rho^{a}\Pi_{i}^{a}\right)$ by $p_{i}$, and according to Theorem \ref{theorem1}, the maximum QFI of $\rho^{b}$ is $F=\sum_{i=1}^{N}\left(\partial_{x}p_{i}\right)^{2}/p_{i}$, which can be achieved if the states in $\{\ket{\psi_{i}}\}$ are orthogonal to each other.
Furthermore, we notice that the maximal QFI of $\rho^{b}$ is equivalent to the CFI of $\rho^{a}$ under projection measurements $\{\Pi_{i}^{a}\}$. 
To achieve $F=F\left(\rho^{a}\right)$, the projection measurements must constitute the optimal measurements of $\rho^{a}$.

In the scenario where the quantum state in auxiliary system is mixed, we have
$O_{i}^{b}\sigma^{b}O_{i}^{b\dagger}=\sum_{j=1}^{M}a_{ij}\ket{\psi_{ij}}\bra{\psi_{ij}}$.
The state after partial trace can be written as 
\begin{equation}
\rho^{b}=\sum_{i=1}^{N}\sum_{j=1}^{M}p_{i}a_{ij}\ket{\psi_{ij}}\bra{\psi_{ij}}.
\end{equation}
The maximal QFI of it is $F=\sum_{i,j}\left(\partial_{x}p_{i}a_{ij}\right)^{2}/p_{i}a_{ij}$, and the coefficient $\{a_{ij}\}$ satisfy $\sum_{j=1}^{M}a_{ij}=1$, thus the above QFI equals to $\sum_{i=1}^{N}\left(\partial_{x}p_{i}\right)^{2}/p_{i}$.
The condition for achieving the maximal value is that $\{\ket{\psi_{ij}}\}$ must be orthogonal, which means the auxiliary system must be prepared in $MN$-dimensional Hilbert space.
To diminish the required dimension of auxiliary system's Hilbert space, the initial state $\sigma^{b}$ should be prepared as pure state. 

Subsequently, we will refine the unitary operator $U$ to further minimize the dimensions required for auxiliary system.
Suppose the $N$ projection operators $\{\Pi_{i}\}$ are divided into $M$ groups, where $M<N$.
Each $i$th group comprises $g_{i}$ projection operators, with the $j$th projection operator in the $i$th group denoted as $\Pi_{j}^{(i)}$. 
This implies constructing $M$ new projectors through a linear combination of $N$ projectors $\{\Pi_{i}\}$.
The new $i$-th projection operator $\Pi_{i}^{'}$, involving $g_{i}$ projector $\Pi_{j}^{(i)}$, can be written as $\Pi_{i}^{'}=\sum_{j=1}^{g_{i}}\Pi_{j}^{(i)}$. 
Using the aforementioned symbolic definitions, we derive the following theorem.
\begin{theorem} \label{theorem2}
Assume a set of $N$ projectors $\{\Pi_{i}\}$ constitutes a measurement of $\rho$ via projection.
For a new set of $M$ projector $\{\Pi'_{i}\}$, defined as $\Pi_{i}'=\sum_{j=1}^{g_{i}}\Pi_{j}$ for $i\in[1,M](M<N)$, with the stipulation that each the $\Pi_{i}$ within $\Pi'_{i}$ is unique and non-repeating.
Denote $\Pi_{j}^{(i)}\rho\Pi_{j}^{(i)}=p_{j}(i)$, if $p_{j}^{(i)}\partial_{x}p_{k}^{(i)}=p_{k}^{(i)}\partial_{x}p_{j}^{(i)}$ holds for $k,l\in[1,g_{i}]$ and every index $i$, then the performance of projectors $\{\Pi_{i}'\}$ is the same as the measurement $\{\Pi_{i}\}$.
\end{theorem}
We consider two different operators, which are $U=\sum_{i=1}^{N}\Pi_{i}^{a}\otimes O_{i}^{b}$ and $\tilde{U}=\sum_{i=1}^{M}\Pi_{i}'\otimes O_{i}^{b}$. From theorem \ref{theorem1}, we can obtain the maximal QFI of state $\rho^{b}$ is $F=\sum_{i=1}^{N}\left(\partial_{x}p_{i}\right)^{2}/p_{i}$. By denoting $\Pi_{j}^{a(i)}\rho^{a}\Pi_{j}^{a(i)}=p_{j}^{(i)}$, the maximal QFI of state $\rho^{b}$ with unitary operator $\tilde{U}$ is
\begin{equation}
\tilde{F}=\sum_{i=1}^{M}\frac{\left(\sum_{j=1}^{g_{i}}\partial_{x}p_{j}^{(i)}\right)^{2}}{\sum_{j=1}^{g_{i}}p_{j}^{(i)}}.
\end{equation} 
The above theorem can be obtained by directly calculating the difference between $F$ and $\tilde{F}$ in Appendix \ref{opt U}.
It is evident that the QFIs, $F$ and $\tilde{F}$, are equivalent to the CFIs under projection measurements $\{\Pi_{i}^{a}\}$ and $\{\Pi_{i}^{a'}\}$, respectively.
If $\{\Pi_{i}^{a}\}$ and $\{\Pi_{i}^{a'}\}$ satisfy the condition mentioned in Theorem \ref{theorem2}, the maximal QFI of $\rho^{b}$ under the unitary operators $U$ and $\tilde{U}$ are the same.
Since the unitary operator $\tilde{U}$ comprises only $M$ operator $\{O_{i}^{b}\}$, the dimension required for auxiliary system is $M$, and $M<N$.
It is apparent that reducing the number of projection operators, provided the aforementioned condition is met, the dimension of auxiliary system can be reduced without influencing the maximal value of QFI.

Experimentally, the probability of the initial system $\rho^{a}$ under a set of projection measurements $\{\Pi_{i}^{a}\}$ is measurable, and the derivative with respect to the parameter $x$ can be derived using the parameter-shift rule~\cite{Mitarai2018,Schuld2019,Li2017,Banchi2021,Cimini2024}, allowing for the assessment of the condition in Theorem \ref{theorem2}.
Then the projection operators $\Pi$ satisfying the above condition are combined linearly to form a new projection operator $\Pi'$. In this way, the number of projection operators is reduced to reduce the required dimension of auxiliary system.

\section{Conclusion}
As a lower bound of QFI, the sub-QFI can be written as Eq. \ref{general sub-QFI}, which is independent of the parameterization method.
Owing to its measurability in experiments, sub-QFI serves as a viable tool for evaluating the QFI of a quantum state.
To streamline experimental complexity, as detailed in the main text, for assessing the QFI of high-dimensional systems.
Initially, we combine the high-dimensional quantum state $\rho^{a}$, which is parameter-dependent, with a parameter-independent quantum state $\sigma^{b}$ in auxiliary system to create a composite system.
By preforming a unitary transformation $U=\sum_{i}\Pi_{i}^{a}\otimes O_{i}^{b}$ and taking the partial trace over system $a$, the parameter information about from the high-dimensional system can be propagated to the auxiliary system. 
Provided that $\sigma^{b}$ is a pure state and all $\{O_{i}^{b}\sigma^{b}O_{i}^{b\dagger}\}$ are mutually orthogonal, the QFI of the auxiliary system reaches its maximum, equivalent to the CFI of the initial state $\rho^{a}(x)$ under the projection operators $\{\Pi_{i}\}$. 
Consequently, to ensure complete propagation of information from state $\rho^{a}(x)$ to the auxiliary system, the projection measurement must be the optimal measurement for the initial state concerning the parameter $x$.
Furthermore,  because the states in $\{O_{i}^{b}\sigma^{b}O_{i}^{b\dagger}\}$ must be orthogonal, the dimension of the auxiliary system is correlate with the number of $\{O_{i}^{b}\}$ and $\{\Pi_{i}^{a}\}$ within the unitary operator $U$.
Assuming there are $g$ projection operators satisfying condition $p_{i}\partial_{x}p_{j}=p_{j}\partial_{x}p_{i}$, for $\forall i,j \in[1,g]$, theses several projection operators can be linearly combined to create a new projection operator, thereby reduce the count of projection operators without compromising the QFI value of the auxiliary system.
Through this approach, the number of projection operators is minimized, thereby reducing the requisite dimension of auxiliary system, achieving the goal of simplifying the complexity of measuring sub-QFI.

\appendix

\section{derivation of sub-QFI} \label{arbitrary sub-QFI}
The sub-QFI can be derived from superfidelity with the following equation
\begin{equation}
F^{(\mathrm{sub})}=8\lim_{\mathrm{d}x\to0}\frac{1-\sqrt{g\left(\rho(x),\rho(x+\mathrm{d}x)\right)}}{\left(\mathrm{d}x\right)^{2}}.
\end{equation}
By performing a second-order Taylor expansion, $\rho(x+\mathrm{d}x)$ can be written as
\begin{equation}
\rho(x+\mathrm{d}x)=\rho(x)+\partial_{x}\rho(x)\mathrm{d}x
+\frac{1}{2}\partial_{x}^{2}\rho(x)\mathrm{d}x^{2}.
\end{equation}
In Eq. (A3), $\partial_{x}^{2}\rho(x)=\partial^{2}\rho/\partial x^{2}$.
Substituting Eq. (A3) into the expressions for $\rho(x)\rho(x+\mathrm{d}x)$ and $\rho^{2}(x+\mathrm{d}x)$ yields
\begin{equation}
\begin{aligned}
&\rho(x)\rho(x+\mathrm{d}x)\\
&=\rho^{2}(x)+\rho(x)\partial_{x}\rho(x)\mathrm{d}x
+\frac{1}{2}\rho(x)\partial_{x}^{2}\rho(x)\mathrm{d}x^{2},
\end{aligned}
\end{equation}
and
\begin{equation}
\begin{aligned}
&\rho^{2}(x+\mathrm{d}x)\\
&=\left(\rho(x)+\partial_{x}\rho(x)\mathrm{d}x
+\frac{1}{2}\partial^{2}_{x}\rho(x)\mathrm{d}x^{2}\right)^{2}\\
&=\rho^{2}(x)+\rho(x)\partial_{x}\rho(x)\mathrm{d}x
+\left(\partial_{x}\rho(x)\right)\rho(x)\mathrm{d}x\\
&+\frac{1}{2}\rho(x)\partial_{x}^{2}\rho(x)\mathrm{d}x^{2}
+\frac{1}{2}\left(\partial_{x}^{2}\rho(x)\right)\rho(x)\mathrm{d}x^{2}\\
&+\left(\partial_{x}\rho(x)\right)^{2}\mathrm{d}x^{2}+\mathcal{O}(\mathrm{d}x^{3}).
\end{aligned}
\end{equation}
Let $\rho$ denote $\rho(x)$, then the expression for $1-\mathrm{Tr}\left(\rho^{2}(x+\mathrm{d}x)\right)$ becomes
\begin{equation}
\begin{aligned}
&1-\mathrm{Tr}\left(\rho^{2}(x+\mathrm{d}x)\right)\\
&=1-\mathrm{Tr}\left(\rho^{2}\right)-2\mathrm{Tr}\left(\rho\partial_{x}\rho\right)\mathrm{d}x\\
&-\left[\mathrm{Tr}\left(\rho\partial^{2}_{x}\rho\right)+\mathrm{Tr}\left((\partial_{x}\rho)^{2}\right)\right]\mathrm{d}x^{2}.
\end{aligned}
\end{equation}
Utilizing Eq. (A5) and the approximation $\sqrt{1-x}\sim 1-x/2-x^{2}/8$ for $x\to0$, 
the expression $\sqrt{(1-\mathrm{Tr}(\rho^{2}))(1-\mathrm{Tr}(\rho(x+\mathrm{d}x)^{2})}$ can be written as
\begin{equation}
\begin{aligned}
&\sqrt{\left(1-\mathrm{Tr}\left(\rho^{2}\right)\right)
\left(1-\mathrm{Tr}\left(\rho(x+\mathrm{d}x)\right)^{2}\right)}\\
&=1-\mathrm{Tr}\left(\rho^{2}\right)-\mathrm{Tr}\left(\rho\partial_{x}\rho\right)\mathrm{d}x
-\frac{1}{2}\left[\mathrm{Tr}\left(\rho\partial_{x}^{2}\rho\right)\right.\\
&\left.+\mathrm{Tr}\left((\partial_{x}\rho)^{2}\right)
+\frac{\left(\mathrm{Tr}\left(\rho\partial_{x}\rho\right)\right)^{2}}
{1-\mathrm{Tr}\left(\rho^{2}\right)}\right]\mathrm{d}x^{2}.
\end{aligned}
\end{equation}
To prevent a divergence denominator, it is assumed that the quantum state under consideration is a mixed state.
Then, superfidelity can be calculated as
\begin{equation}
\begin{aligned}
&g\left(\rho(x),\rho(x+\mathrm{d}x)\right)\\
&=1-\frac{1}{2}\left[\mathrm{Tr}\left((\partial_{x}\rho(x))^{2}\right)
+\frac{\left(\mathrm{Tr}\left(\rho(x)\partial_{x}\rho(x)\right)\right)^{2}}
{1-\mathrm{Tr}\left(\rho^{2}(x)\right)}\right]\mathrm{d}x^{2}.
\end{aligned}
\end{equation}
As $\mathrm{d}x\to 1$, the expression $1-\sqrt{g\left(\rho(x),\rho(x+\mathrm{d}x)\right)}$ approaches
\begin{equation}
\begin{aligned}
&1-\sqrt{g\left(\rho(x),\rho(x+\mathrm{d}x)\right)}\\
&=\frac{1}{8}\left[2\mathrm{Tr}\left((\partial_{x}\rho(x))^{2}\right)
+\frac{1}{2}\frac{\left(\partial_{x}\mathrm{Tr}\left(\rho^{2}(x)\right)\right)^{2}}
{1-\mathrm{Tr}\left(\rho^{2}(x)\right)}\right]\mathrm{d}x^{2}.
\end{aligned}
\end{equation}
Combine Eq. (A1) and Eq. (A8), sub-QFI can be written as
\begin{equation}
F^{(\mathrm{sub})}
=2\mathrm{Tr}\left((\partial_{x}\rho(x))^{2}\right)
+\frac{1}{2}\frac{\left(\partial_{x}\mathrm{Tr}\left(\rho^{2}(x)\right)\right)^{2}}
{1-\mathrm{Tr}\left(\rho^{2}(x)\right)}.
\end{equation}
If the probe state is parameterized by a unitary transformation, 
that is $\rho=e^{-ixH}\rho_{0}e^{ixH}$, 
$\mathrm{Tr}\left(\left(\partial_{x}\rho(x)\right)^{2}\right)$ and $\mathrm{Tr}\left(\rho^{2}(x)\right)$
can be written as
\begin{equation}
\begin{aligned}
&\mathrm{Tr}\left(\left(\partial_{x}\rho\right)^{2}\right)=2\mathrm{Tr}\left(\rho^{2}H^{2}\right)-2\mathrm{Tr}\left(\rho H\rho H\right),
\end{aligned}
\end{equation}
and
\begin{equation}
\mathrm{Tr}\left(\rho^{2}(x)\right)=\mathrm{Tr}\left(e^{-ixH}\rho^{2}e^{ixH}\right)=\mathrm{Tr}\left(\rho^{2}\right).
\end{equation}
Given that $\partial_{x}\mathrm{Tr}(\rho^{2})=0$, the sub-QFI for unitary parameterization becomes
\begin{equation}
F^{(\mathrm{sub})}=4\mathrm{Tr}\left(\rho^{2}H^{2}\right)-4\mathrm{Tr}\left(\rho H\rho H\right).
\end{equation}

For a pure state, where $\mathrm{Tr}(\rho^{2})=1$, the superfidelity simplifies to
$g(\rho_{1},\rho_{2})=\mathrm{Tr}(\rho_{1}\rho_{2})$, and
\begin{equation}
\mathrm{Tr}\left(\rho\rho(x+\mathrm{d}x)\right)=\mathrm{Tr}(\rho^{2})
+\frac{1}{2}\mathrm{Tr}\left(\rho\partial_{x}^{2}\rho\right)\mathrm{d}x^{2}.
\end{equation}
With the identity $\mathrm{Tr}\left[\left(\partial_{x}\rho\right)^{2}\right]=\mathrm{Tr}\left(\rho\partial_{x}^{2}\rho\right)$, the sub-QFI for a pure state is
\begin{equation}
F^{(\mathrm{sub})}=2\mathrm{Tr}\left[\left(\partial_{x}\rho\right)^{2}\right],
\end{equation}
which equals to QFI of it.

Let $s(X)=\sum_{i\neq j}\lambda_{i}(X)\lambda_{j}(X)$ denote the sum of products of distinct eigenvalues of $X$, the superfidelity and the square of fidelity are then expressed as
\begin{eqnarray}
&&g(\rho_{1},\rho_{2})=\mathrm{Tr}(\rho_{1}\rho_{2})+2\sqrt{s(\rho_{1})s(\rho_{2})},\\
&&f^{2}\left(\rho_{1},\rho_{2}\right)=\mathrm{Tr}(\rho_{1}\rho_{2})+2s\left(\sqrt{\sqrt{\rho_{1}}\rho_{2}\sqrt{\rho_{1}}}\right).
\end{eqnarray}
For single-qubit states $\rho_{1}$ and $\rho_{2}$, where $s(\rho_{i})=\lambda_{1}(\rho_{i})\lambda_{2}(\rho_{i})$, it is observed that $s\left(\sqrt{\sqrt{\rho_{1}}\rho_{2}\sqrt{\rho_{1}}}\right)=s\left(\sqrt{\sqrt{\rho_{1}}\rho_{2}\sqrt{\rho_{1}}}\right)$, indicating that superfidelity equals to the fidelity for single qubits. 
Consequently, the sub-QFI for both pure states and single-qubit states is found to be equivalent to their QFI.

\section{Optimization of single-qubit auxiliary system} \label{opt single-qubit}
When the quantum state in auxiliary system is a single qubit, the quantum state $\rho^{b}$ to be measured can be expressed using Eq. (9) from the main text.
Because $\rho_{1}$ and $\rho_{2}$ are density operators obtained by $\sigma^{b}$ undergoing different unitary operators, the norm of Bloch vectors $|\vec{r}_{1}|=|\vec{r}_{2}|$, and denoted by $r$.
The Bloch vector of $\rho^{b}$ is given by $\vec{r}=p_{1}\vec{r}_{1}+p_{2}\vec{r}_{2}$, and can be represented in spherical coordinates as follows,
\begin{equation}
\left\{
\begin{array}{l}
r_{x}=r(p_{1}\sin\theta_{1}\cos\phi_{1}+p_{2}\sin\theta_{2}\cos\phi_{2}), \\
r_{y}=r(p_{1}\sin\theta_{1}\sin\phi_{1}+p_{2}\sin\theta_{2}\sin\phi_{2}), \\
r_{z}=r(p_{1}\cos\theta_{1}+p_{2}\cos\theta_{2}).
\end{array}
\right.
\end{equation}

The QFI in terms of the Bloch representation is given by~\cite{Zhong2013}
\begin{equation}
F=\left(\partial_{x}\vec{r}\right)^{2}+\frac{\left(\vec{r}\cdot\partial_{x}\vec{r}\right)^{2}}{1-\left|\vec{r}\right|^{2}}.
\end{equation}
To derive its explicit form, we require the components of the gradient vector $\partial_{x}\vec{r}$, which can be expressed as 
\begin{equation}
\left\{
\begin{array}{l}
\partial_{x}r_{x}=r(\partial_{x}p_{1}\sin\theta_{1}\cos\phi_{1}+\partial_{x}p_{2}\sin\theta_{2}\cos\phi_{2}), \\
\partial_{x}r_{y}=r(\partial_{x}p_{1}\sin\theta_{1}\sin\phi_{1}+\partial_{x}p_{2}\sin\theta_{2}\sin\phi_{2}), \\
\partial_{x}r_{z}=r(\partial_{x}p_{1}\cos\theta_{1}+\partial_{x}p_{2}\cos\theta_{2}).
\end{array}
\right.
\end{equation}
Furthermore, we introduce a quantity $\chi(\theta,\phi)$, related to the angles in the spherical coordinate system, deined as $\chi(\theta,\phi)=\sin\theta_{1}\sin\theta_{2}\cos(\phi_{1}-\phi_{2})+\cos\theta_{1}\cos\theta_{2}-1$, then each part of the QFI with respect to the Bloch vector can be calculated as
\begin{eqnarray}
&&\left(\partial_{x}\vec{r}\right)^{2}=-2r^{2}\left(\partial_{x}p_{1}\right)^{2}\chi(\theta,\phi), \\
&&\left|\vec{r}\right|^{2}=r^{2}\left(1+2p_{1}p_{2}\chi(\theta,\phi)\right), \\
&&\vec{r}\cdot\partial_{x}\vec{r}=r^{2}\left(p_{1}\partial_{x}p_{2}+p_{2}\partial_{x}p_{1}\right)\chi(\theta,\phi).
\end{eqnarray}
In the following, for the convenience of writing, denote $\chi=\chi(\theta,\phi)$. 
It is straightforward to observe that permissible the range for $\chi$ is $\chi\in[-2,0]$.
Employing the aforementioned equations, the expression for QFI is deduced to be
\begin{equation}
F=\frac{\left[-2r^{2}(1-r^{2})\chi+r^{4}\chi^{2}\right]\left(\partial_{x}p_{1}\right)^{2}}{1-r^{2}-2r^{2}p_{1}p_{2}\chi}.
\end{equation}
The expression for $F$ clearly indicates a dependence on both $\chi$ and $r$.
To maximize $F$, we initially optimize the value of $\chi$.
By differentiating $F$ with respect to $\chi$ and solving for zero derivative, we determine the extreme point $\chi_{\pm}$ of $F$ to be
\begin{equation}
\chi_{\pm}=\frac{(1-r^{2})\left(1\pm\sqrt{1-4p_{1}p_{2}}\right)}{2r^{2}p_{1}p_{2}}.
\end{equation}
Given that $1\pm\sqrt{1-4p_{1}p_{2}}>0$ and considering the range of $\chi$ is $[-2,0]$, the only existing extreme point only occurs at $r=1$ with $\chi=0$, yielding $F|_{\chi=0}=0$.
At the boundary of $\chi$, specifically when $\chi=-2$, we have $F|_{\chi=-2}=4r^{2}(\partial_{x}p_{1})^{2}/(1-r^{2}+4r^{2}p_{1}p_{2})>0$, and this value is greater than that as $\chi=0$. 
Consequently, to maximize $F$, $\chi$ should be set to $-2$.
And then we optimize $F|_{\chi=-2}$ with respect to $r$.
By examining the derivative of $F|_{\chi=-2}$ with respect to $r$, we conclude that $F|_{\chi=-2}$ increases monotonically. 
Hence, the maximum value of $F|_{\chi=-2}$ occurs at $r=1$.
In summary, the maximum value of $F$ is achieved when $\chi=-2$ and $r=1$.
With $r=1$ indicating that the quantum state $\sigma^{b}$ is a pure state, while $\chi=-2$ signifying that $\rho_{1}$ and $\rho_{2}$ are orthogonal, and the maximum value of $F$ is
\begin{equation}
F_{\mathrm{max}}=\frac{\left(\partial_{x}p_{1}\right)^{2}}{p_{1}p_{2}}=\frac{\left(\partial_{x}p_{1}\right)^{2}}{p_{1}}+\frac{\left(\partial_{x}p_{1}\right)^{2}}{p_{2}}.
\end{equation}

\section{Prove of Theorem 1.} \label{orthogonal bases}
We consider quantum state with the form $\rho=\sum_{i}p_{i}\ket{\psi_{i}}\bra{\psi_{i}}$, assuming that the set $\{\ket{\psi_{i}}\}$ is linearly independent, thus forming a basis.
In Ref.~\cite{Genoni2019}, Genoni and Tufarelli proposed a method for calculating QFI under non-orthogonal bases, which we utilize in the subsequent calculations.
Specifically, let the density operator and its derivative with parameter are expressed as $\rho=\sum_{ij}R_{ij}\ket{\psi_{i}}\bra{\psi_{j}}$ and $\partial_{x}\rho=\sum_{ij}D_{ij}\ket{\psi_{i}}\bra{\psi_{j}}$, respectively.
The QFI of $\rho$ is
\begin{equation}
F=\mathrm{Re}\left[\mathrm{Tr}(SLSLSR)\right].
\end{equation}
In equation (23), matrix $L$ represents the symmetric logarithmic derivative (SLD) operator under these bases and elements of matrix $S$ are given by $S_{ij}=\braket{\psi_{i}|\psi_{j}}$.
These matrices satisfy the relation $2D=RSL+LSR$. 
Employing the above method, we proceed to proof Theorem \ref{theorem1}.
\begin{proof}
For $\rho=\sum_{i}p_{i}\ket{\psi_{i}}\bra{\psi_{i}}$, with $\{p_{i}\}$ depending on the parameter and $\{\ket{\psi_{i}}\}$ being independent of it, the derivative of $\rho$ with respect to $x$ is $\partial_{x}\rho=\sum_{i}\partial_{x}p_{i}\ket{\psi_{i}}\bra{\psi_{i}}$.
Thus, the matrices $R$ and $D$ are diagonal, that is, $R_{ij}=p_{i}\delta_{ij}$ and $D_{ij}=\partial_{x}p_{i}\delta_{ij}$.
To calculate the QFI, we utilize the transformed basis $\{\ket{\phi_{i}}=\sqrt{p_{i}}\ket{\psi_{i}}\}$ in place of the original $\{\ket{\psi_{i}}\}$ to construct the density operator.
Under this new basis, density operator $\rho$ and its derivative $\partial_{x}\rho$ are expressed as
\begin{equation}
\rho=\sum_{i}p_{i}\ket{\psi_{i}}\bra{\psi_{i}}=\sum_{i}\ket{\phi_{i}}\bra{\phi_{i}},
\end{equation}
and 
\begin{equation}
\partial_{x}\rho=\sum_{i}\partial_{x}p_{i}\ket{\psi_{i}}\bra{\psi_{i}}=\sum_{i}\frac{\partial_{x}p_{i}}{p_{i}}\ket{\phi_{i}}\bra{\phi_{i}}.
\end{equation}
Hence, $R$ equals the identity matrix $\mathbb{I}$, and the elements of Gram matrix $S$ are $S_{ij}=\sqrt{p_{i}p_{j}}\braket{\psi_{i}|\psi_{j}}$.
Subsequently, we establish an identity
\begin{equation}
\begin{aligned}
4\mathrm{Tr}(DSD)&=\mathrm{Tr}\left[(SL+LS)S(SL+LS)\right]\\
&=3\mathrm{Tr}(SLSLS)+\mathrm{Tr}(LSSSL).
\end{aligned}
\end{equation}
$\mathrm{Tr}(DSD)$ can be calculated as
\begin{equation}
\begin{aligned}
\mathrm{Tr}(DSD)
&=\sum_{i}\sum_{kl}\frac{\partial_{x}p_{i}}{p_{i}}\delta_{ik}\sqrt{p_{k}p_{l}}\braket{\psi_{i}|\psi_{k}}
\frac{\partial_{x}p_{i}}{p_{i}}\delta_{il}\\
&=\sum_{i}\frac{(\partial_{x}p_{i})^{2}}{p_{i}}.
\end{aligned}
\end{equation}
Let $M=SL$, from the equation $2D=RSL+LSR$, we derive
\begin{equation}
\left\{
\begin{array}{cc}
\mathrm{Re}(M_{ii})=\frac{\partial_{x}p_{i}}{p_{i}} , \\
M_{ij}+M_{ji}^{\ast}=0, & i\neq j .
\end{array}
\right.
\end{equation}
Thus, $[MM^{\dagger}]_{ij}$ and $[M^{\dagger}M]_{ij}$ satisfy
\begin{equation}
\begin{aligned}
\left[MM^{\dagger}\right]_{ij}&=\left[M^{\dagger}M\right]_{ij}+M_{ii}^{\ast}M_{ji}^{\ast}+M_{ij}M_{jj}\\
&+M_{ii}M_{ji}^{\ast}+M_{ij}M_{jj}^{\ast}\\
&=\left[M^{\dagger}M\right]_{ij}+2\mathrm{Re}(M_{ii})M_{ji}^{\ast}+2\mathrm{Re}(M_{jj})M_{ij}\\
&=\left[M^{\dagger}M\right]_{ij}+2\left(\frac{\partial_{x}p_{j}}{p_{j}}-\frac{\partial_{x}p_{i}}{p_{i}}\right)M_{ij}.
\end{aligned}
\end{equation}
By using equation (30), we calculate $\mathrm{Tr}(LSSSL)$ as
\begin{equation}
\begin{aligned}
&\mathrm{Tr}(LSSSL)=\mathrm{Tr}(SMM^{\dagger})\\
&=\mathrm{Tr}(SLSLS)
+2\sum_{i,j}S_{ij}\left(\frac{\partial_{x}p_{i}}{p_{i}}-\frac{\partial_{x}p_{j}}{p_{j}}\right)M_{ji}
\end{aligned}
\end{equation}
By combining equation (26) and (31), we obtain
\begin{equation}
\begin{aligned}
&4\mathrm{Tr}(DSD)\\
&=3\mathrm{Tr}(SLSLS)+\mathrm{Tr}(LSSSL)\\
&=4\mathrm{Tr}(SLSLS)
+2\sum_{i,j}S_{ij}\left(\frac{\partial_{x}p_{i}}{p_{i}}-\frac{\partial_{x}p_{j}}{p_{j}}\right)M_{ji}.
\end{aligned}
\end{equation}
The QFI is
\begin{equation}
\begin{aligned}
F&=\mathrm{Re}\left[\mathrm{Tr}(SLSLS)\right]\\
&=\mathrm{Tr}(DSD)-\frac{1}{2}\mathrm{Re}\left[\sum_{i,j}S_{ij}\left(\frac{\partial_{x}p_{i}}{p_{i}}
-\frac{\partial_{x}p_{j}}{p_{j}}\right)M_{ji}\right]\\
&=\sum_{i}\frac{(\partial_{x}p_{i})^{2}}{p_{i}}
-\frac{1}{2}\mathrm{Re}\left[\sum_{i,j}S_{ij}\left(\frac{\partial_{x}p_{i}}{p_{i}}
-\frac{\partial_{x}p_{j}}{p_{j}}\right)M_{ji}\right]\\
&=\sum_{i}\frac{(\partial_{x}p_{i})^{2}}{p_{i}}
-\frac{1}{2}\sum_{i,j}S_{ij}\left(\frac{\partial_{x}p_{i}}{p_{i}}
-\frac{\partial_{x}p_{j}}{p_{j}}\right)\left(SL\right)_{ji}.
\end{aligned}
\end{equation}

In Ref.~\cite{Alipour2015}, Alipour and Rezakhani derive an inequality of quantum state of the form $\rho=\sum_{i}p_{i}\rho_{i}$:
\begin{equation}
F\left(\sum_{i}p_{i}\rho_{i}\right)\le \sum_{i}p_{i}F(\rho_{i})+I\left(\{p_{i}\}\right).
\end{equation}
Here, $I(\{p_{i}\})$ is the CFI of the probability distribution $\{p_{i}\}$.
Assuming $\rho_{i}$ are parameter-independent, implying $F(\rho_{i})=0$, we have
\begin{equation}
F\left(\sum_{i}p_{i}\rho_{i}\right)\le \sum_{i}\frac{(\partial_{x}p_{i})^{2}}{p_{i}}.
\end{equation}
To maximize the QFI of $\rho=\sum_{i}p_{i}\ket{\psi}\bra{\psi}$, the second term
in equation (B14) must vanish, that is
\begin{equation}
\sum_{i,j}S_{ij}\left(\frac{\partial_{x}p_{i}}{p_{i}}
-\frac{\partial_{x}p_{j}}{p_{j}}\right)(SL)_{ji} = 0.
\end{equation}
For the condition $\partial_{x}p_{i}/p_{i}=\partial_{x}p_{j}/p_{j}$ establish for every $i\neq j$, it must be that
\begin{equation}
p_{j}\partial_{x}p_{i}=p_{i}\partial_{x}p_{j}, i\neq j.
\end{equation}
Summing of equation (B17) and applying the normalization $\sum_{i}p_{i}=1$ along with $\sum_{i}\partial_{x}p_{i}=0$, we deduce that
\begin{equation}
\begin{aligned}
&\left(\sum_{i}p_{i}\right)\partial_{x}p_{j}
=p_{j}\left(\sum_{i}\partial_{x}p_{i}\right)\\
&\Rightarrow \partial_{x}p_{j}=0.
\end{aligned}
\end{equation}
Consequently, the identity $\partial_{x}p_{i}/p_{i}=\partial_{x}p_{j}/p_{j}$ for all $i\neq j$ is only satisfied if $\partial_{x}p_{j}=0$ for any $j$, a condition that cannot be met.
Provided that the basis state are mutually orthogonal, such that $S_{ij}=p_{i}\delta_{ij}$,
\begin{equation}
\begin{aligned}
&\sum_{i,j}S_{ij}\left(\frac{\partial_{x}p_{i}}{p_{i}}
-\frac{\partial_{x}p_{j}}{p_{j}}\right)(SL)_{ji}\\
&=\sum_{ij}p_{i}\delta_{ij}\left(\frac{\partial_{x}p_{i}}{p_{i}}-\frac{\partial_{x}p_{j}}{p_{j}}\right)(SL)_{ki}=0.
\end{aligned}
\end{equation}
In conclusion, the QFI of $\rho=\sum_{i}p_{i}\ket{\psi_{i}}\bra{\psi_{i}}$ is maximized when the states $\{\ket{\psi_{i}}\}$ are orthogonal, yielding $F_{\mathrm{max}}=\sum_{i}(\partial_{x}p_{i})^{2}/p_{i}$.
\end{proof}

In our approach mentioned in the main text, the state resulting from the partial trace is given by $\rho^{b}=\sum_{i}p_{i}\rho_{i}$.
When the individual states $\rho_{i}$ are pure state, we have $\rho^{b}=\sum_{i}p_{i}\ket{\psi_{i}}\bra{\psi_{i}}$.
Consequently, the maximal QFI for $\rho^{b}$ is expressed as $F_{\mathrm{max}}=\sum_{i}(\partial_{x}p_{i})^{2}/p_{i}$.
In the scenario where the states $\{\rho_{i}\}$ are mixed, they can be represented as $\rho_{i}=\sum_{j}a_{ij}\ket{\psi_{ij}}\bra{\psi_{ij}}$, and the coefficients $\{a_{ij}\}$ satisfy $\sum_{j}a_{ij}=1$.
Thus, the quantum state $\rho^{b}$ can be written as $\rho=\sum_{i,j}p_{i}a_{ij}\ket{\psi_{ij}}\bra{\psi_{ij}}$.
The corresponding maximum QFI is given by
\begin{equation}
F_{\mathrm{max}}=\sum_{i,j}\frac{\left(\partial_{x}p_{i}a_{ij}\right)^{2}}{p_{i}a_{ij}}
=\sum_{i}\frac{\left(\partial_{x}p_{i}\right)^{2}}{p_{i}}.
\end{equation}
This is equivalent to the maximum QFI found in the case of pure states.
The condition for achieving the maximum value is that all states $\{\ket{\psi_{ij}}\}$ are orthogonal.
To ensure that the set of states $\{\ket{\psi_{i}}\}$ are orthogonal to one another,  the dimension of Hilbert space must be at least equal to the number of different quantum states in $\{\ket{\psi_{i}}\}$.
It can be deduced that lower-dimensional space is sufficient when the state $\sigma^{b}$ in auxiliary system consists of pure state, as opposed to mixed state.
Therefore, we opt for a pure-state auxiliary system to reduce the complexity of the measurement process.

\section{Prove of Theorem 2.} \label{opt U}
Consider two sets of projection measurements $\{\Pi_{i}\}$ and $\{\Pi_{i}'\}$. 
The $M$ projectors $\{\Pi_{i}'\}$ are constructed from $\{\Pi_{i}'\}$ through linear combinations.
Let $\Pi_{j}^{(i)}$ denote the $j$th constituent projector in $i$th composite projector $\Pi_{i}'$, which comprises $g_{i}$ projectors $\Pi_{j}^{(i)} (j\in[1,g_{i}])$, and $\Pi_{i}'=\sum_{j=1}^{g_{i}}\Pi_{j}^{(i)}$.
For $\{\Pi_{i}'\}$ to qualify as a set of projection operators, each $\Pi_{i}'$ contain unique $\{\Pi_{j}^{(i)}\}$ without repetition.
Using the above notation, we will prove Theorem \ref{theorem2} in the following.
\begin{proof}
The above two projection measurements $\{\Pi_{i}\}$ and $\{\Pi_{i}'\}$ yielding two CFI $I_{1}$ and $I_{2}$ for a quantum state $\rho$.
Denote $p_{i}=\mathrm{Tr}(\Pi_{i}\rho\Pi_{i})$, $I_{1}$ and $I_{2}$ can be written as
\begin{eqnarray}
I_{1}&=&\sum_{i=1}^{N}\frac{\left(\partial_{x}p_{i}\right)^{2}}{p_{i}}, \\
I_{2}&=&\sum_{i=1}^{M}\frac{\left(\sum_{j=1}^{g_{i}}\partial_{x}p_{j}^{(i)}\right)^{2}}{\sum_{j=1}^{g_{i}}p_{j}^{(i)}}.
\end{eqnarray}
By directly calculating the difference between $I_{1}$ and $I_{2}$,
\begin{eqnarray}
&&I_{1}-I_{2} \nonumber \\
&&=\sum_{i=1}^{N}\frac{\left(\partial_{x}p_{i}\right)^{2}}{p_{i}}
-\sum_{i=1}^{M}\frac{\left(\sum_{j=1}^{g_{i}}\partial_{x}p_{j}^{(i)}\right)^{2}}{\sum_{j=1}^{g_{i}}p_{j}^{(i)}}
\nonumber \\
&&=\sum_{i=1}^{M}\sum_{j=1}^{g_{i}}\frac{\left(\partial_{x}p_{j}^{(i)}\right)^{2}}{p_{j}^{(i)}}
-\sum_{i=1}^{M}\frac{\left(\sum_{j=1}^{g_{i}}\partial_{x}p_{j}^{(i)}\right)^{2}}{\sum_{j=1}^{g_{i}}p_{j}^{(i)}}
\nonumber \\
&&=\sum_{i=1}^{M}\sum_{j,l=1 \atop (j\neq l)}^{g_{i}}
\left[\frac{p_{l}^{(i)}\left(\partial_{x}p_{j}^{(i)}\right)^{2}}{p_{j}^{(i)}\sum_{k=1}^{g_{i}}p_{k}^{(i)}}
-\frac{\partial_{x}p_{j}^{(i)}\partial_{x}p_{l}^{(i)}}{p_{l}^{(i)}\sum_{k=1}^{g_{i}}p_{k}^{(i)}}\right]
\nonumber \\
&&=\sum_{i=1}^{M}\sum_{j,l=1 \atop (j<l)}^{g_{i}}
\frac{\left(p_{l}^{(i)}\partial_{x}p_{j}^{(i)}-p_{j}^{(i)}\partial_{x}p_{l}^{(i)}\right)^{2}}
{p_{j}^{(i)}p_{l}^{(i)}\sum_{k=1}^{g_{i}}p_{k}^{(i)}} \ge 0.
\end{eqnarray}
Consequently, reducing the number of projection operators through linear combination will reduce the CFI, implying $I_{1}\ge I_{2}$.
Specifically, when for every $i$, the condition $p_{l}^{(i)}\partial_{x}p_{j}^{(i)}=p_{j}^{(i)}\partial_{x}p_{l}^{(i)}$ is met for $j,l \in [1,g_{i}]$, the numerator in the above equation is identical to zero, that is $F_{1}=F_{2}$.
\end{proof}

Consider the unitary operator $U_{1}$ constructed from $N$ projectors $\{\Pi_{i}\}$, defined as $U_{1}=\sum_{i=1}^{N}\Pi_{i}\otimes O_{i}^{b}$.
From this unitary operator, we can obtain $\rho^{b}_{1}=\sum_{i=1}^{N}p_{i}\rho_{i}$ in auxiliary system, whose maximal QFI is equal to $I_{1}$.
Using the projectors from $U_{1}$, we form a new set of projectors $\{\Pi_{i}^{'}\}$ through linear combination, which consists $M (M<N)$ projectors, and $\Pi_{i}^{'}=\sum_{j=1}^{g_{i}}\Pi_{j}^{(i)}$.
With the set $\{\Pi_{i}^{'}\}$, we can construct a new unitary operator $U_{2}$,
\begin{equation}
U_2=\sum_{i=1}^{M}\Pi_{i}^{'}\otimes O_{i}^{b}
=\sum_{i=1}^{M}\left(\sum_{j=1}^{g_{i}}\Pi_{j}^{(i)}\right)\otimes O_{i}^{b}.
\end{equation}
For this scenario, the resulting partial trace yields the density operator $\rho^{b}_{2}=\sum_{i=1}^{M}\left(\sum_{j=1}^{g_{i}}p_{j}^{(i)}\right)\rho_{i}$. 
And the maximal QFI of $\rho_{2}^{b}$ is $I_{2}$.
It can be seen that satisfying the condition in Theorem \ref{theorem2} ensures that the QFI of the final results obtained from these two unitary operators are equal.
Furthermore, the dimension of the auxiliary system is dictated by the number of projection operators.
Consequently, identifying projection operators that meet the aforementioned condition and combining them linearly to form new projectors in a revised unitary operator allows for a reduction in the required dimension of auxiliary system while preserving the QFI value.

\end{document}